\documentclass[sigconf]{acmart}
\PassOptionsToPackage{table}{xcolor} % must be before any package that might load xcolor
\usepackage{xcolor}
\usepackage{colortbl} % provides \rowcolor, \columncolor, etc.
\definecolor{lightgreen}{RGB}{220,255,220}

\usepackage[english]{babel}
\usepackage{blindtext}

\usepackage{tabularx}
\usepackage{booktabs}
\usepackage{adjustbox}
\usepackage{graphicx} % Keep this single instance
\usepackage{multirow} % Keep this single instance

\usepackage{array}
\usepackage{diagbox}
\usepackage{threeparttable}
\usepackage{amsmath}
\usepackage{amsfonts}
\usepackage{amssymb}
\usepackage{wasysym}  % For special symbols

\usepackage{algorithm}
\usepackage{algpseudocode}
\usepackage{hyperref}
\usepackage{subcaption} 
\usepackage{pifont}
\usepackage{caption}
\usepackage{tikz}

\newcommand*\circledm[1]{\tikz[baseline=(char.base)]{
            \node[shape=circle, inner sep=1pt, fill=red] (char) {\textcolor{white}{#1}};}}
\usepackage{lipsum} % For dummy text (optional)
\usepackage{pifont}  % For the filled and empty circles

\usepackage{amsmath}

\AtBeginDocument{%
  }
\newcommand{\NetGuard}{\texttt{NetGuard }}
\newcommand{\fullcircle}{\ding{108}}  % Full black circle
\setcopyright{acmlicensed}
\copyrightyear{}
\acmYear{}
\acmDOI{}
\acmConference[Conference KDD'25]{}{}{}
\acmISBN{XXX}
\begin{document}

\title{Generative Active Adaptation for Drifting and Imbalanced Network Intrusion Detection}
\author{Ragini Gupta$^{1}$, Shinan Liu$^{2}$, Ruixiao Zhang$^{1}$, Xinyue Hu$^{1}$, Xiaoyang Wang$^{1}$, Hadjer Benkraouda$^{1}$, Pranav Kommaraju$^{1}$, Phuong Cao$^{1}$, Nick Feamster$^{2}$, Klara Nahrstedt$^{1}$}
\affiliation{%
$^{1}$ University of Illinois Urbana-Champaign  \city{Urbana} \state{IL} \country{USA},
$^{2}$ University of Chicago \city{Chicago} \state{IL} \country{USA}
\\
\{raginig2, ruixiao, xinyue21, xw28, hadjerb2, pvk4, pcao3, klara\}@illinois.edu, \{shinanliu, feamster\}@uchicago.edu
}
\renewcommand{\shortauthors}{}

\begin{abstract}
 Machine learning has shown promise in network intrusion detection systems, yet its performance often degrades due to concept drift and imbalanced data. These challenges are compounded by the labor-intensive process of labeling network traffic, especially when dealing with evolving and rare attack types, which makes preparing the right data for adaptation difficult. To address these issues, we propose a generative active adaptation framework that minimizes labeling effort while enhancing model robustness. Our approach employs density-aware dataset prior selection to identify the most informative samples for annotation, and leverages deep generative models to conditionally synthesize diverse samples, thereby augmenting the training set and mitigating the effects of concept drift. We evaluate our end-to-end framework \NetGuard on both simulated IDS data and a real-world ISP dataset, demonstrating significant improvements in intrusion detection performance. Our method boosts the overall F1-score from 0.60 (without adaptation) to 0.86. Rare attacks such as Infiltration, Web Attack, and FTP-BruteForce, which originally achieved F1 scores of 0.001, 0.04, and 0.00, improve to 0.30, 0.50, and 0.71, respectively, with generative active adaptation in the CIC-IDS 2018 dataset. Our framework effectively enhances rare attack detection while reducing labeling costs, making it a scalable and practical solution for intrusion detection.
\end{abstract}


\maketitle 
\section{Introduction}
\vspace{-0.12cm}
Machine learning (ML) has significantly improved network intrusion detection systems (NIDS), surpassing traditional signature-based methods. Unlike manually curated signatures, ML-based NIDS can detect unknown attacks, especially with unsupervised techniques \cite{r1}. Various approaches — supervised \cite{aids1,aids2,aids3, aids4, aids5, aids6,liu2024serveflow,wan2024cato}, unsupervised \cite{uaids1,uaids2,uaids3, uaids4}, and semi-supervised \cite{semi1, semi2} — have been explored. Supervised models (e.g., CNN, MLP, XGBoost) achieve high accuracy but require large labeled datasets. Unsupervised methods (e.g., K-means, Kitsune \cite{uaids3}) use unlabeled data but often with lower accuracy. Semi-supervised techniques reduce labeling needs but may perform inconsistently on smaller datasets. Despite these challenges, ML- and DL-based NIDS have demonstrated over 95\% detection accuracy in controlled tests \cite{ML1,r1}.

Despite their promise, current ML-driven NIDS frequently assume that training and testing data follow a static distribution. In reality, the network environment is subject to \textbf{concept drift}\cite{cade}, where the statistical behaviors of both benign and malicious traffic evolve, thus the models trained on the original data fail to capture the changes. These changes arise from zero-day attacks, shifting user patterns, and instant updates to network services. In our research, a model trained on a well-known intrusion dataset \cite{cic18} for one year saw its F1 score fall from 0.99 to 0.61 when tested on data from the following year, even though the types of attacks remained largely the same.
Similar observations were made using the real-world UGR’16 \cite{ugr16} dataset, which contains anonymized NetFlow traces collected over four months in a Tier-3 ISP. This dataset integrates realistic attack scenarios with labeled traffic. When a classifier trained on one month of data (July) was tested on another month's data (August), the F1 score dropped from 0.96 to 0.687, emphasizing the impact of temporal shifts on the NIDS model.

\noindent \textbf{State of the Art.} Traditional approaches to handle concept drift include retraining the model by combining new data with old data or retraining from scratch on the new data only \cite{tesseract,r2}. However, these methods can be computationally expensive and risk \emph{catastrophic forgetting}, where the model’s performance on previously learned data degrades as new information is introduced \cite{catforget}. Other advanced techniques for addressing concept drift in network intrusion detection rely on active learning to adapt models to evolving distributions \cite{r3,r5,r6}. While numerous tools exist for collecting and capturing network data, labeling remains a significant challenge due to the time, expertise, and resources required. Active learning addresses this bottleneck by prioritizing the labeling of the most informative samples, enabling models to update effectively and maintain performance in dynamic environments. The adaptation of active learning methods in network intrusion detection mainly stems from different sampling strategies: \textit{uncertainty, diversity, density}, or a combination of these. Uncertainty-based methods prioritize samples with the greatest predictive uncertainty, as seen in OWAD \cite{owad}. Diversity-based approaches, often combined with uncertainty sampling (e.g., ENiDrift \cite{enidrift}), seek to maximize the variability of labeled instances to improve generalization. Density-based methods, like CADE \cite{cade} and LEAF \cite{leaf}, leverage embedding distances to identify structurally representative samples in feature space. We found that uncertainty-based methods struggle to select informative attack samples when the majority of the benign samples are drifting, leading to biased models that miss rare attacks. Additionally, they require a high labeling budget, as many approaches still demand substantial annotations to sustain performance.

\begin{figure*}[!ht]
    \centering
    %----------- Subfigure (a) -----------
    \begin{subfigure}[b]{0.49\textwidth}
        \includegraphics[width=\textwidth]{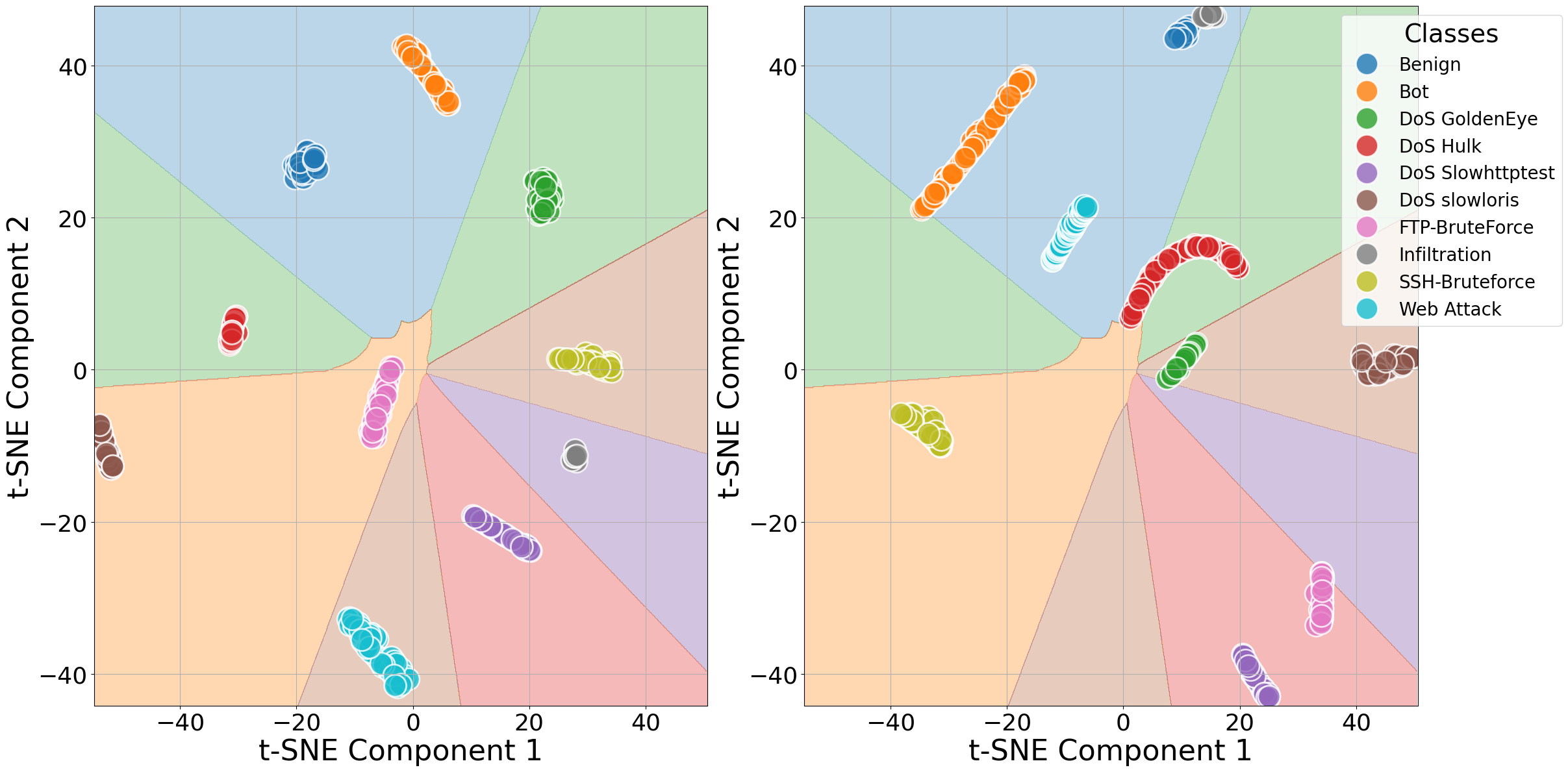}
        \caption{Common attacks in CIC 2017 (left) and CIC 2018 (right)}
        \label{fig:subfig1}
    \end{subfigure}
    \hfill
    %----------- Subfigure (b) -----------
    \begin{subfigure}[b]{0.49\textwidth}
        \includegraphics[width=\textwidth]{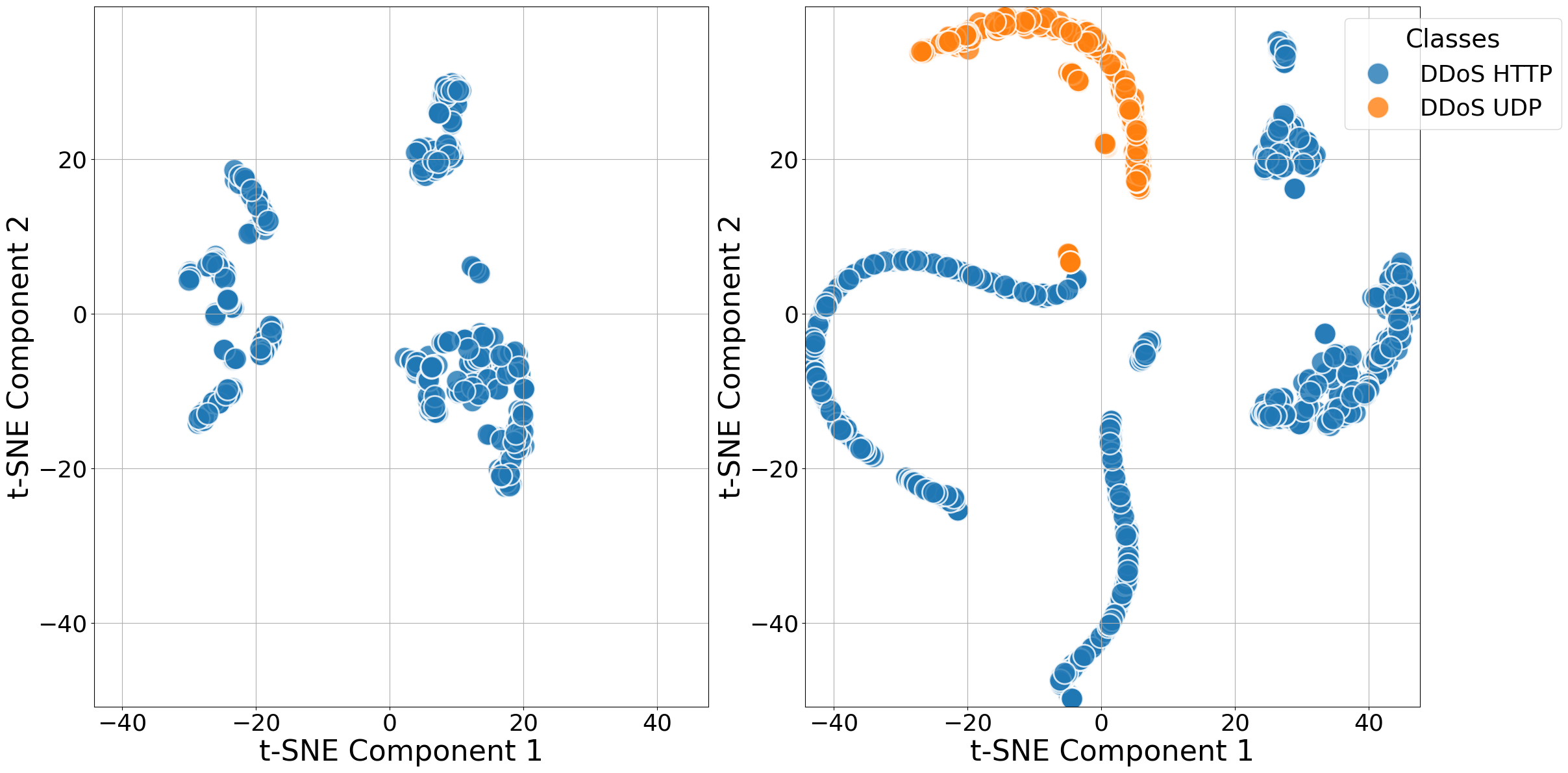}
        \caption{New variants of DDoS attacks in  CIC 2017 (left) and CIC 2018 (right)}
        \label{fig:subfig2}
    \end{subfigure}
    \caption{\small t-SNE visualization of embedding space for both CIC 2017/2018 datasets}
    \label{fig:cicDist}
\end{figure*}

\noindent \textbf{Challenges}. Building robust, learning-enabled NIDS is impeded by two major challenges: \underline{Concept Drift} and \underline{Class Imbalance}.
(1) Network traffic patterns are highly dynamic, leading to significant concept drift. First, the evolution of normal traffic—driven by the proliferation of IoT devices, 5G/6G networks, and shifts in user behavior—continuously redefines benign activities~\cite{r1}. Second, attackers adapt their existing strategies, refining techniques to bypass detection~\cite{mitre}. Third, entirely new attack types can emerge suddenly, rendering previous models less effective   \cite{divar,divar2} (we illustrated this in Fig.\ref{fig:cicDist} (b) with new variants of DDoS attack in CIC 2018). Additionally, overlapping embeddings between benign and malicious traffic further blur the distinction between classes, complicating the task of accurate classification~\cite{uaids3}.
(2) Class imbalance poses another critical challenge. NIDS datasets often exhibit a skewed distribution where benign samples vastly outnumber malicious ones, particularly for rare or emerging attack types~\cite{malware1}. This scarcity of minority class examples degrades the performance of ML models, as they become biased toward the majority class. Moreover, the low presence of attack samples hampers active adaptation; uncertainty-based active learning methods struggle to select informative samples when critical minority examples are underrepresented. This limitation not only reduces overall detection accuracy but also slows the system's ability to adapt quickly to evolving threats.

\noindent \textbf{Our Contributions}. We propose an end-to-end framework, called \NetGuard, that employs a density-aware dataset prior selection strategy along with conditional generative data augmentation to address concept drift and class imbalance between training (or source) and testing (or target) network intrusion detection datasets. The drifted and rare attacks are underrepresented in NIDS datasets. Na\"ively applying actively learned samples for retraining further exacerbates the problem of data scarcity. This is particularly problematic for training NIDS models, where reduced training data can lead to overfitting and inflated performance metrics for detection. To address this, there are two potential approaches: \textit{first}, acquiring more samples, which essentially contradicts our goal for optimizing labeling cost, and second, synthetically generating more data points. The \textit{second} approach takes the assumption that the synthesized dataset is representative of the original distribution. Thus, \NetGuard identifies a dataset prior and conditionally synthesizes more diverse data, thus enhancing the robustness and generalization of NIDS models when the training and testing datasets differ due to temporal evolution or their origination from different network architectures or domains. We make the following contributions.
\noindent \textit{A. Density and Informativeness}. Our algorithm employs a practical scenario with a labeled training dataset (\(X_L\)) and an unlabeled testing dataset (\(X_{UL}\)). The algorithm uses a distribution-aware GMM scheme to identify samples in \(X_{UL}\) that exhibit locally sparse density and differ significantly from the distribution of the training dataset \(X_L\). This ensures that the selected dataset priors are highly informative, representing areas of the feature space where the data distribution shifts, improving the sample quality for labeling.

\noindent \textit{B. Diversity via Data Augmentation}. To counter the challenges of class imbalance, our framework leverages deep generative models for conditional data augmentation on the dataset priors. By employing few-shot generation techniques, we synthesize diverse samples in underrepresented areas of the feature space. This augmentation not only creates a more balanced dataset but also improves the model's ability to generalize to new and evolving attack types.
\vspace{-0.4cm}
\section{Background}
\vspace{-0.1cm}
\subsection{Related Works}
\noindent \textbf{Concept Drift Adaptations.} Concept drift is a key challenge in real-world NIDS due to evolving protocols, devices, user behavior, and new attack vectors. Foundational Surveys~\cite{surveycd1,surveycd2,surveycd3,surveycd4} emphasize its prevalence and negative impact on detection accuracy. However, most existing NIDS solutions handle either drift or class imbalance, not both. Table 1 compares representative works against four features that are critical for real-world deployment: concept drift handling, imbalanced data mitigation, multi-class support, and labeling cost. For instance, CADE~\cite{cade} and ENiDrift~\cite{enidrift} both address concept drift, but only ENiDrift manages imbalanced data effectively; neither is designed to handle multi-class scenarios or minimize labeling requirements. ReCDA~\cite{recda} adapts to drift using self-supervision but still incurs high labeling cost and struggles with rare attacks. LEAF~\cite{leaf} focuses on drift in regression tasks for cellular networks, but lacks support for supervised, imbalanced, multi-class intrusion detection. In contrast, \NetGuard addresses all four dimensions—adapting to drift, managing imbalance, supporting multi-class attacks, and reducing labeling overhead—making it more practical for real-world NIDS. In contrast, \NetGuard framework aims to encompass all these dimensions—continuously adapting to drift, managing imbalance, supporting multi-class attacks, and reducing labeling overhead—thus offering a solution that aligns with real-world needs. Recent zero-shot and meta-learning approaches~\cite{cdnips1,cdnips2,cdnips3,cdnips4} adapt quickly in unsupervised domains like time series or images, but they are not suited for supervised NIDS. These methods detect deviations from “normal” without labels, but fail when class labels persist under shift. Their reliance on meta-distributions also limits their use in open-world scenarios where novel attacks lack prior structure. 

\noindent \textbf{Active Learning}. Active Learning (AL) \cite{ALSurvey} reduces the labeling burden by selectively annotating the most informative samples, assuming an accurate labeling agent. Traditional uncertainty-based acquisition functions~\cite{liu2023amir}, such as entropy, margin sampling, and least confidence, struggle with imbalanced datasets, as minority class samples are often misclassified with high confidence and thus not selected for labeling \cite{AL90}. Existing AL approaches for network intrusion detection \cite{aladin, aladin2} focus on querying uncertain or rare/new attacks but fail to address concept drift, where prior class distributions may change. Other AL approaches, such as Coreset \cite{coreset}, focus on data diversity to improve coverage across different class distributions. However, most of the AL methods either rely on model predictions or select samples without considering their relationship to previously seen data. This lack of comparison between new unlabeled data and previously trained data can lead to suboptimal sample selection. Recent methods address fast or zero-shot anomaly detection under distributional shift, primarily in unsupervised domains such as multivariate time series or image data~ \cite{cdnips1,cdnips2,cdnips3,cdnips4}. These approaches leverage meta-learning or prototype transfer for hypernetwork tuning to enable rapid adaptation, but they are not tailored for supervised NIDS with tasks with class-labeled and imbalanced attack classes. While effective at detecting deviations from the "normal" class without labels, types. These methods can identify samples that deviate from "normal" without requiring class labels, but they cannot prioritize high-impact samples when class labels remain fixed under distributional shift. They also rely on access to meta-distributions or related tasks during training \cite{cdnips1,cdnips2,cdnips3}, which enables few-shot generalization but limits applicability in open-world settings, real-time network environments where novel unseen attacks emerge without prior task structure.

\noindent \textbf{Generative AI for Network Data Synthesis}
Rare but critical attacks are underrepresented in network traffic, leading to high false negatives in learning-enabled NIDS. Additionally, publicly available network datasets are often obsolete, failing to capture evolving traffic patterns, while privacy concerns prevent ISPs and businesses from sharing real-world data. This scarcity limits the training of robust ML-based models for intrusion detection and traffic analysis. Recent advances in generative AI, including GANs \cite{doppleGan,netshare}, diffusion models \cite{netdiffusion,jiang2023generative}, GPT-based models \cite{trafficgpt}, state-space models \cite{chu2024feasibility}, hybrid Transformer-GPT architectures \cite{rtf}, and VAEs \cite{tvae,tvae2}, have shown promise in synthesizing high-fidelity network traces to expand the feature space and improve model generalization for network analysis. However, most of these methods focus on zero-shot augmentation, which requires domain expertise or class-conditional generation using all available samples, without accounting for evolving distributions during test time. Other augmentation works, such as RAPIER~\cite{rapier} addresses label noise via GAN-based augmentation under static distributions, a problem orthogonal to our framework where we assume labels are correct under dynamic data distributions.
To the best of our knowledge, prior works in building robust NIDS address either concept drift within a single dataset~\cite{cade, recda} or apply generic data augmentation for class imbalance \cite{netshare}. In contrast, we propose a novel approach that integrates state-of-the-art generative models with conditional generation, guided by drift-aware prior selection. This enables targeted synthesis of minority-class traffic under evolving distributions, offering a practical solution for adaptive NIDS in real-world settings. Our work is among the first to examine cross-domain drift using feature-aligned datasets (CIC-IDS 2017 and 2018), offering a realistic test of model robustness across deployments. For datasets like UGR'16 with different features, we instead evaluate temporal drift within a single domain.
\vspace{-0.35cm}
\begin{table}[h!]
    \centering
    \scriptsize % Reduce font size
    \caption{\small Comparison of representative related works}
    \renewcommand{\arraystretch}{1} % Slightly reduced row height
    \setlength{\tabcolsep}{1pt} % Reduce column spacing
    \begin{tabular}{l c c c c c c}
        \toprule
        \textbf{Features} & \textbf{CADE \cite{cade}} & \textbf{ENiDrift \cite{enidrift}} & \textbf{ReCDA \cite{recda}}  & \textbf{LEAF \cite{leaf}} & \textbf{CLUE \cite{clue}} & \textbf{NetGuard} \\
        \midrule
        Concept drift   & \fullcircle & \fullcircle & \fullcircle & \fullcircle & \fullcircle  & \fullcircle \\
        Imbalanced data & \Circle     & \fullcircle & \Circle     & \Circle     &\Circle        & \fullcircle \\
        Multi-class     & \Circle     & \Circle     & \Circle     & \fullcircle & \fullcircle  & \fullcircle \\
        Labeling cost   & \Circle     & \Circle     & \Circle     & \Circle     & \fullcircle       & \fullcircle \\
        Sampling        & Density     & Uncert.+Div. & Density & Density   & Uncert.+Div. & Dens.+Div. \\
        \bottomrule
    \end{tabular} \\
    \footnotesize (\fullcircle = true, \Circle = false)
\end{table}
\vspace{-0.2cm}

\vspace{-0.4cm}
\subsection{Motivation}
\vspace{-0.15cm}
We derive the motivation for \NetGuard by evaluating how existing active adaptation approaches perform on CIC-IDS 2017/2018. Fig. \ref{fig:cicDist} illustrates the latent space distribution of benign and attack classes across the two datasets. In Fig. \ref{fig:cicDist}(a), the left graph shows the class distribution in CIC-IDS 2017, while the right graph represents the same classes in CIC-IDS 2018, focusing on common attacks. As evident, while the MLP decision boundaries trained on 2017 CIC data (left graph) remain unchanged (right graph), there is a significant class distribution between the two datasets, confirming the presence of concept drift. The figure demonstrates that the same attack classes occupy different regions in the feature space across datasets, indicating distributional changes between two domains. Similarly, Fig. \ref{fig:cicDist}(b) shows DDoS HTTP latent space distribution in CIC-IDS 2017 (left graph), while new DDoS variants emerge in CIC-IDS 2018 (right right). Table 2 presents the sample strength of different attack classes in CIC-IDS 2017 and 2018, highlighting severe class imbalance. The Earth Mover's Distance (EMD) or Wassertein-1 distance column measures the distance between two probability distributions between the same classes in the two datasets, with higher EMD indicating significant drift. When we apply different AL sampling approaches on CIC-IDS 2018, Uncertainty-based AL \cite{uncertainty} fails to sample from minority classes like Bot, FTP-BruteForce, and Web Attack. Similarly, Uncertainty + Diversity (CLUE) \cite{clue} improves sample selection slightly but still fails to prioritize rare classes. On the other hand, our {\tt NetGuard}'s density-aware sample selection strategy selects more samples from classes with high EMD (such as Infiltration, DDoS, DoS Slowloris), ensuring that samples representative of evolving distribution shifts are included in retraining. However, it alone does not resolve class imbalance, as underrepresented attacks remain scarce even with optimized sampling. To address this, we adapt it as an approach to select priors for state-of-the-art deep generative models, enabling data augmentation to improve the detection of rare attacks. 
\vspace{-0.35cm}
\begin{table}[h!]
    \small
    \setlength{\tabcolsep}{2pt}
    \renewcommand{\arraystretch}{1.3}
    \centering
    \begin{adjustbox}{max width=\linewidth}
    \begin{tabular}{lccccccc}
        \toprule
        \textbf{Class} &
        \shortstack{\textbf{CIC 2017} \\ \textbf{\#Samples}} &
        \shortstack{\textbf{CIC 2018} \\ \textbf{\#Samples}} &
        \shortstack{\textbf{Norm.} \\ \textbf{EMD}} &
\shortstack{\textbf{Uncertainty} \\ {                                 } } & \shortstack{\textbf{Diversity} \\ \textbf{(CoreSet)}} &
        \shortstack{\textbf{Uncert. +} \\ \textbf{Diversity (CLUE)}} &
        \shortstack{\textbf{\NetGuard} \\ \textbf{PS}} \\
        \midrule
        Benign         & 37937 & 23494 & 1.0000 & 63  & 366 & 159 & 249 \\
        Bot            & 1956  & 9682  & 0.3691 & 0   & 24  & 36  & 19  \\
        DDoS           & 11555 & 11018 & 0.5170 & 44  & 57  & 66  & 129 \\
        DoS GoldenEye  & 9078  & 6850  & 0.1216 & 578 & 53  & 100 & 30  \\
        DoS Hulk       & 12131 & 10112 & 0.1066 & 5   & 14  & 102 & 26  \\
        Slowhttptest   & 5499  & 9038  & 0.2159 & 0   & 21  & 65  & 53  \\
        Slowloris      & 5796  & 1813  & 0.2478 & 37  & 40  & 29  & 61  \\
        FTP-BruteForce & 7935  & 9329  & 0.0616 & 0   & 19  & 50  & 15  \\
        Infiltration   & 36    & 9140  & 0.6161 & 21  & 165 & 61  & 153 \\
        SSH-BruteForce & 5897  & 9302  & 0.2062 & 15  & 36  & 131 & 51  \\
        Web Attack     & 2180  & 151   & 0.1073 & 36  & 4   & 0   & 26  \\
        \bottomrule
    \end{tabular}
    \end{adjustbox}
    \caption{\small Class distributions for CIC datasets, normalized EMD, and samples selected per strategy. PS = Prior Selection.}
    \label{tab:sampling}
\end{table}
\vspace{-0.2cm}
\section{Methodology}\vspace{-0.12cm}\subsection{Design Overview}
Figure 2 illustrates the end-to-end workflow of {\tt NetGuard}, which consists of two key modules: density-aware dataset prior selection and conditional data augmentation for minority attack classes. The process begins with training a deep learning-based classifier on the original labeled training dataset ($X_L$). Once trained, the classifier is tested on a new, unlabeled testing dataset ($X_{UL}$), where it predicts labels for the unseen data. If the performance metrics degrade, the prior selection module is triggered to enhance model performance. In the density-aware dataset prior selection stage, a Gaussian Mixture Model (GMM) is fitted to both $X_L$ and $X_{UL}$. Using the learned distributional information and an informativeness metric, the most representative and informed samples $X_P$ are selected within a given budget. These selected samples are then queried by a human annotator or an oracle for labeling. The annotator continuously expands the original training dataset by adding the newly labeled samples, which helps improve the model’s generalization capability. Additionally, this step addresses the issue of dataset diversity, ensuring that low-distribution (minority) attack classes are well represented. To further enhance model robustness, minority attack classes with less than 5\% distribution are fed into the data augmentation module. Unlike generic augmentation, it performs conditional generation synthesizing samples conditioned on the actively selected, high-informative priors from $X_{UL}$. This ensures that synthetic samples reflect the current test-time distribution and preserve attack semantics. The generated samples are integrated into the training corpus, producing a more balanced dataset and strengthening the model's robustness to detect underrepresented and evolving attacks. Thus, \NetGuard introduces a novel conditional data augmentation strategy, anchored on actively selected samples, to improve detection under concept drift and class imbalance.

\vspace{-3mm}
\begin{figure}[h!] % Single-column figure
    \centering
    \includegraphics[width=0.5\textwidth]{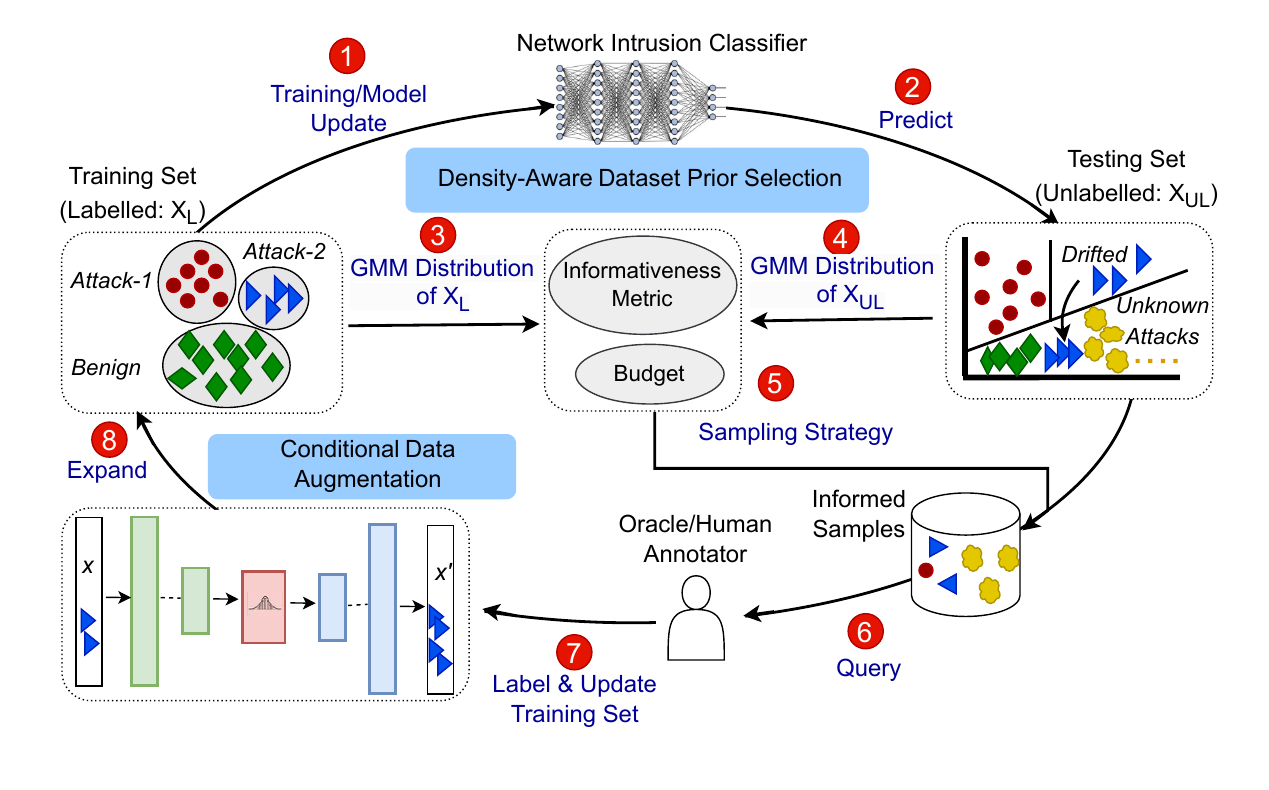} % Adjusted for single-column
    \caption{\small Closed-loop learning framework of \NetGuard}
    \label{fig:wide_figure}
\end{figure}
% \vspace{-3mm}

\subsection{Density-aware Dataset Prior Selection}

\subsubsection{Problem Formulation}
Let \( X_L \) and \( X_{UL} \) be the training and testing datasets, respectively, where \( X_L \) is labeled and \( X_{UL} \) is unlabeled batch. Both datasets contain \( C \) or more classes, including various attack types extracted from different network architectures. Due to potential differences in network architectures, service types, and data collection methods between the two datasets, their feature spaces \( X_L \) and \( X_{UL} \) and probability distributions \( P(X_L) \) and \( P(X_{UL}) \) may differ significantly. This variation results in the concept drift problem, where a deep learning based classification model \(f\)  trained on \( X_L \) using a learning algorithm to minimize a loss function \( L \) may experience performance degradation when applied to \( X_{UL} \), leading to \( \text{Performance}(f, X_L) \neq \text{Performance}(f, X_{UL}) \).

Therefore, to solve this problem and given the distributional discrepancy between \( X_L \) and \( X_{UL} \), we devise a new informativeness metric \( I_{Score}\) within the prior selection framework by drawing inspiration from the well-known unsupervised clustering model, Gaussian Mixture Model (GMM), a powerful way to probabilistically represent latent distribution with its parameter set formalized as $\psi \sim (\pi, \mu, \sigma^2)$,
where $\psi$ encapsulates the mixture weights ($\pi$) component means ($\mu$), and covariances ($\sigma^2$) of the GMM. These parameters are learned using the Expectation-Maximization (EM) algorithm. Using the learned distributional information allows us to estimate the log-likelihood ($p(z)$) of unlabeled samples \( x_{UL} \)  with respect to both the labeled dataset \( X_L \) and the unlabeled dataset \( X_{UL} \) which is then embedded into the selection criterion to compute \( I_{Score}\). Samples with high \( I_{Score}\) values are prioritized within the Prior labeling budget \( P \)  for retraining the model. 

\vspace{-0.2cm}
\subsubsection{Workflow}
In each iteration, the classifier selects the most informative samples from unlabeled data \( X_{UL} \) based on a predefined informativeness criterion, queries an oracle for labels, and expands \( X_L \). This process continues until the labeling budget \( P \) is exhausted or a stopping condition is met.

\noindent \textbf{Initial Model Training \circledm{1}} In AL, a classifier \( f \) is trained in iterative rounds \( t = 0,1,\dots,T \), starting with a DNN-based model initially trained on the labeled dataset \( X_L \).

\noindent \textbf{Prediction on Unlabeled Dataset  \circledm{2}} The trained model makes label predictions on the unlabeled dataset \( X_{UL} \).

\noindent \textbf{Fit GMMs to \( X_L \) and \( X_{UL} \)  \circledm{3}  \circledm{4}}  To characterize the latent distributions of the labeled and unlabeled datasets, we fit a GMM to the labeled dataset \( X_L \) to obtain the parameters \( \psi_L = (\pi_L, \mu_L, \sigma_L^2) \). Similarly, another GMM is fitted to the unlabeled dataset \( X_{UL} \) to obtain the parameters \( \psi_{UL} = (\pi_{UL}, \mu_{UL}, \sigma_{UL}^2) \).

\noindent \textbf{Compute Log-Likelihoods \circledm{5}} The intuition behind sample selection is that those samples should be chosen which are dissimilar to \( X_L \), but at the same time they are the best representatives of \( X_{UL} \). Subsequently, using the learned GMMs, the log-likelihood is computed for each unlabeled sample \( x_{UL} \). First, we compute its log-likelihood under the GMM of the unlabeled dataset, resulting in: \( p(z_{UL}; \psi_{UL}) = \sum_{j=1}^{K} \pi_{UL,j} \mathcal{N}(z_{UL} | \mu_{UL,j}, \sigma_{UL,j}^2) \). Then, the log-likelihood of the same sample under the GMM of the labeled dataset is computed as: \(
p(z_{UL}; \psi_L) = \sum_{j=1}^{K} \pi_{L,j} \mathcal{N}(z_{UL} | \mu_{L,j}, \sigma_{L,j}^2) \). Note that the log-likelihood $p(z_{UL}; \psi_{UL})$ is calculated as it represents the probability that a given data of interest is generated by the GMM characterized by $\psi$.\\
\textbf{Calculate Informativeness Score \circledm{5}} The informativeness score for a data point $x_{UL}$ is defined as the difference between its log-likelihood under the unlabeled dataset's GMM and its log-likelihood under the labeled dataset's GMM. Notationally, this is \( I_{\text{Score}}(x_{UL}) = p(x_{UL}; \psi_{UL}) - p(x_{UL}; \psi_L) \). This score quantifies the most valuable data points $x_{UL}$ as they help the model reduce distributional gaps between the training and testing datasets.
\noindent \textbf{Prior Selection Strategy  \circledm{5}} The acquisition function $A(x)$ is defined as the set of all informativeness scores $I_{\text{score}}$ values for $x \in X_{UL}$:
\( A(x) = \{I_{\text{score}}(x) \mid x \in X_{UL}\} \). To select informative samples, the scores are sorted in descending order, and the top $P$ samples are chosen based on the available labeling budget $P$. This ensures that the batch $X_P \subseteq X_{UL}$ consists of the most informative samples for labeling. It is worth mentioning that by sorting samples in descending order of their informativeness score, we prioritize those with high $ I_{\text{Score}}$, thereby selecting samples that both \textbf{improve coverage} and \textbf{maximize informativeness}. Samples that improve coverage are those with high $ p(x_{UL}; \psi_{UL}) $ that are representative of the unlabeled dataset and help the model learn from regions of the feature space that are underrepresented in $ X_L $. Meanwhile, samples that maximize informativeness are those with low $ p(x_{UL}; \psi_L) ,$ which are essentially dissimilar to the labeled dataset $X_L$ and likely come from regions where the model's predictions are uncertain.  
% Thus, this density-aware sample selection strategy allows the model to select samples from low-density regions in the labeled dataset 

\noindent \textbf{Obtain Ground Truth Labels  \circledm{6}} Send the selected subset $X_P$ to the human annotator (or an oracle) to obtain ground-truth labels. 

\noindent \textbf{Update Labeled/Unlabeled Sets \circledm{7}}
Add these newly labeled samples \((X_P, Y_P)\) to the labeled training set \(X_L\). Remove \(X_P\) from unlabeled dataset \(X_{UL}\).
\vspace{-0.42cm}
\subsection{Diversity-aware Conditional Data Augmentation}After the prior selection step, the sampled points are included in the original labeled dataset (\(X_L\)), which remains imbalanced with under-represented classes, referred to as the minority classes. A threshold of 5\% is used to identify minority classes. The goal is to synthesize new samples using deep generative techniques \circledm{8} such that (a) the sample count of each minority class matches that of the majority class, ensuring class balance, (b) the generated samples maintain representativeness within a single class by capturing the underlying distribution of each minority class, but they should be different enough to prevent redundancy (near duplicate samples) and subsequent model overfitting and (c) the synthetic samples generated for different classes should remain distinct from each other with clear separation between classes in feature space.

\noindent \textbf{Problem Formulation}. Let \( X_P \) be the set of samples obtained from the density-aware prior selection module using a given labeling budget \( P \). After being labeled by the Oracle, each sample in \( X_P \) corresponds to a row in the tabular dataset, where each row is a \( d \)-dimensional feature vector. The corresponding class labels are denoted as \( Y_P \). Assume this batch contains \( C_P \) different classes. These actively learned labeled samples \( (X_P, Y_P) \) are added to the existing labeled training dataset \( X_L \), forming an updated dataset \( X'_L \) with labels \( Y'_L \). The expanded dataset spans a set of classes \( C' \), which includes previously seen classes from \( X_L \) and any newly introduced classes from \( X_P \). Despite the expansion, the dataset may still be imbalanced, with some classes having fewer than 5\% of the total samples. Let \( C'_{\min} \) represent the set of such minority classes, and let \( X'_{\text{min}} \) be the subset of training samples that belong to these minority classes, $X'_{\text{min}} = \{(x'_i, y'_i) \mid y'_i \in C'_{\min} \}$. Let \( G \) be a generative model trained on the minority classes \( X'_{\text{min}} \).

\noindent \textbf{Deep Generative Training.} The goal of \( G \) is to learn the true joint distribution on the different metadata and measurement features of minority classes, represented as \( P_{\text{min}}(x') \). The model \( G \), parameterized by \( \theta \), learns an approximation \( P_{Gen}(x'; \theta) \) on the true distribution. \( G \) is trained by minimizing the loss function \( L_G \) that measures the divergence between the actual minority class distribution \( P_{\text{min}}(x') \) and the generated distribution \( P_{Gen}(x'; \theta) \): \( \theta^* = \arg\min_{\theta} L_G(P_{\text{min}}(x'), P_{Gen}(x'; \theta)) \), where \( L_G \) is the divergence measure, such as Kullback-Leibler (KL) divergence. Once trained, \( G \) generates synthetic samples by sampling from different regions of its learned distribution. 

Specifically, the generative model achieves diversity by interpolating between different states of metadata conditioned on measurement based features, thereby creating variations within each minority class. Metadata features, such as flow ID, protocol type (like HTTP, HTTPs, FTP, SSH, and email protocols), source/destination port, source/destination IP address, timestamp, TCP flag count (for flags like ACK, FIN, SYN, URG, RST) is set represent discrete states (e.g., \textit{s1,s2,s3} for protocol type). Measurement features, such as packet size, packet header length, flow duration, and flow packets/second, flow bytes/second, represent continuous attributes. The generative model captures the joint distribution of these features, enabling it to interpolate between different states of metadata and measurement features to create diverse synthetic samples.

For example, consider a minority class where the metadata feature $m_1$ (\textit{protocol type}) has states such as $s_1$ (HTTP), $s_2$ (FTP),  $s_3$ (SSH) and the metadata feature $m_2$ (\textit{source IP range}) has states $s_4$ (range A), $s_5$ (range B).   The generative model can interpolate between these states, such as generating samples with combinations like $(s_1,s_5), (s_2,s_7)$ or $(s_3,s_9)$ while conditioning on measurement features like  $z_1$ (\textit{packet size}) and $z_2$ (\textit{flow duration}).  Formally, we can represent it as $x'_{\text{Gen}} = G\bigl(s_{i}, s_{j} \;\big|\; z_{1}, z_{2}\bigr)$, where $x'_{\text{Gen}}$ is a synthetic sample generated from the minority class by combining metadata states $s_{i}, s_{j}$ with measurement features $z_{1}, z_{2}$. This interpolation creates variations in the synthetic samples, ensuring that the generated data captures a wide range of scenarios within the minority class. In practice, the choice of generative models implicitly addresses diversity in synthetic generation by their distinct mechanisms for learning and sampling from the joint distribution of features. The augmented dataset is then formed by combining these synthetic samples with the original training data, improving class balance while preserving diversity. 

\noindent \textbf{Post-Processing}. After generating synthetic samples, a post-processing step is applied to remove low-quality samples that may resemble benign data. This is done using a binary logistic regression classifier trained to distinguish between benign and non-benign samples. The classifier is trained on the original labeled dataset \( X_L \) and then used to evaluate the generated synthetic samples. Since all synthetic samples belong to the non-benign class, only those with a low probability of being benign are retained. This filtering step ensures that synthetic samples do not closely resemble majority-class benign samples, improving the overall quality of the augmented dataset. By applying this post-processing step, we effectively clean the synthetic data and remove samples that may not contribute to improving minority class representation.

\noindent \textbf{Validation and Classification}. We evaluate the fidelity of synthetic samples using two approaches: (i) by computing the Wasserstein-2 (W2) distance to ensure the overall structure and spacing of synthetic data points are preserved, and (ii) by assessing the performance of downstream classification tasks. %For (i), we compare the original minority class samples from the actively learned set with the augmented samples of the same class. For (ii), we use classification as the downstream task for which we train a deep neural network (DNN) and a tree-based classifier to evaluate classification performance. %If the synthetic samples effectively represent the minority class while also introducing diversity, they should enhance the classifier's ability to generalize. This improvement should be reflected in better performance on the held-out test set, while the training set is a mixture of original and augmented data for a given minority class.

To achieve this, after the augmentation step, the final training set is constructed by combining all minority samples (both synthetic and original) with the majority class samples from the original training dataset $X_L$ as well as the actively selected samples $X_P$. We train both the classifiers on the expanded dataset. For evaluation, we use the future testing dataset $X_{UL}$ such that the samples used during the prior selection phase i.e., $X_P$  are excluded from $X_{UL}$. 
\vspace{-0.2cm}
\section{Evaluation}
\vspace{-0.14cm}
We evaluate \NetGuard end-to-end performance (including prior selection and data augmentation) based on F1-score improvement under drift, FPR, FNR, Accuracy, and rare class F1 improvement. We also compare the performance scores and labeling cost of \NetGuard with state-of-the-art concept drift adaptation. We also analyze whether augmenting rare classes enhances classification while measuring the fidelity of synthetic samples. For synthetic data augmentation, we conduct experiments on an NVIDIA A100 GPU in Google Cloud and benchmark its utilization and latency overhead. Due to space limitations, we provide the computational cost analysis of \NetGuard modules in the Appendix Table 9.
\vspace{-0.3cm}
\subsection{Datasets and Model Selection}
To evaluate \NetGuard, we use two popular NIDS datasets— CIC-IDS (2017/2018) from Canadian 
Institute of Cybersecurity ~\cite{cic18} and real-world UGR'16 network flows from an ISP in Spain ~\cite{ugr16} capturing temporal drift (time-based shifts) and spatial drift (cross-domain/infrastructure differences). For CIC-IDS, we test bidirectional scenarios (1) Train on 2017 → Test on 2018 capturing both temporal and spatial drifts \footnote{CIC 2017 uses small enterprise LAN with 4 attack, 10 victim machines vs. CIC 2018's large hybrid LAN/cloud setup with 10 attack, > 400 victim machines, NAT/DHCP}), and the reverse scenario (2) Train on 2018 → Test on 2017 that mainly reflects spatial drift despite the reversed temporal order. It is worth mentioning that CIC-IDS is commonly used for studying network drift and imbalance~\cite{infocom,cade,malware1}, and similarly, UGR'16 serves as a benchmark for distribution shift and traffic generation ~\cite{netshare,neuripsugr16,ugr16shift}, making these well-suited for \NetGuard. Adversarial drift is out of scope, as generating protocol-compliant attack traffic requires specialized testbeds~\cite{advdrift}. We use a standard 70/30 train-test split, with the training set as source domain and test set as target domain to evaluate drift. We employ an MLP (two 100-neuron hidden layers) for CIC-IDS to model non-linear traffic patterns ~\cite{classifierMLP,classifierxgbtree,classifiermlp2,mlpmine} and XGBoost for UGR'16 ~\cite{classifierxgbtreegood,classifierxgbgood2, classifierxgbtree2} to handle class imbalance, using 100k subsamples per dataset with Web Attack, Infiltration, and FTP-BruteForce (CIC-IDS) and Botnet/Spam (UGR'16) as minority classes. Our model choices align with recent comparisons of neural and non-neural classifiers in prior cybersecurity applications ~\cite{malware1}. To To address imbalance, we generate high-quality synthetic samples at 3×/10× ratios using deep generative models: (a) TVAE (Tabular Variational Autoencoder) \cite{tvae} and (b) RealTabFormer (RTF) \cite{rtf}. TVAE leverages a variational autoencoder framework for structured tabular data, 
while RTF, a transformer-based model, employs sequence-to-sequence (Seq2Seq) 
modeling to capture feature dependencies and structured patterns in tabular data.
\vspace{-0.2in}
\begin{table}[h!]
\begingroup
\renewcommand{\baselinestretch}{0.55}\footnotesize
\setlength{\tabcolsep}{1pt} % tighter column spacing
\renewcommand{\arraystretch}{1} % shrink row spacing
\centering
\caption{\small MLP performance on CIC 2017/2018 datasets}
\label{tab:combined_results}
\resizebox{\columnwidth}{!}{%
\begin{tabular}{p{1cm}l|cccc|cccc}
\toprule
& & \multicolumn{4}{c|}{\textbf{Train: 2017 \(\rightarrow\) Test: 2018}} & \multicolumn{4}{c}{\textbf{Train: 2018 \(\rightarrow\) Test: 2017}} \\
\textbf{Method} & \textbf{Label} & \textbf{F1(\%)} & \textbf{FNR} & \textbf{FPR} & \textbf{Acc.(\%)} & \textbf{F1(\%)} & \textbf{FNR} & \textbf{FPR} & \textbf{Acc.(\%)} \\
\midrule
\textbf{None} & 0\% & 60.62 & 0.3062 & 0.5495 & 60.62 & 58.28 & 0.4925 & 0.3803 & 58.28 \\
\midrule
\multirow{3}{*}{\textbf{Coreset}}    
& 0.1\% & 61.74 & 0.2929 & 0.5453 & 70.71 & 80.55 & 0.1739 & 0.3734 & 82.61 \\
& 0.5\% & 66.49 & 0.2769 & 0.3894 & 72.31 & 84.06 & 0.1421 & 0.3353 & 85.79 \\
& 1\%   & 63.34 & 0.3201 & 0.4427 & 68.0  & 87.28 & 0.1065 & 0.2860 & 89.35 \\
\midrule
\multirow{3}{*}{\textbf{Uncert.}}    
& 0.1\% & 66.70 & 0.2681 & 0.4424 & 73.19 & 73.14 & 0.3267 & 0.2142 & 67.33 \\
& 0.5\% & 67.73 & 0.2387 & 0.4985 & 76.13 & 82.25 & 0.1902 & 0.2125 & 80.98 \\
& 1\%   & 74.18 & 0.2087 & 0.4009 & 79.13 & 84.41 & 0.1545 & 0.2716 & 84.55 \\
\midrule
\multirow{3}{*}{\textbf{CLUE}}    
& 0.1\% & 62.46 & 0.2923 & 0.5292 & 70.77 & 91.63 & 0.0864 & 0.1406 & 91.6 \\
& 0.5\% & 75.24 & 0.2230 & 0.3431 & 77.71 & 94.38 & 0.0610 & 0.0686 & 93.90 \\
& 1\%   & 79.71 & 0.1661 & 0.3443 & 83.39 & 95.98 & 0.0398 & 0.0564 & 96.01 \\
\midrule
\multirow{3}{*}{\begin{tabular}{c}\textbf{CADE+}\\\textbf{Retrain}\end{tabular}}   
& 0.1\% & 33.03 & 0.6278 & 0.0659 & 43.22 & 35.38 & 0.5875 & 0.0546 & 49.10 \\
& 0.5\% & 24.95 & 0.7128 & 0.0723 & 37.95 & 32.0  & 0.5754 & 0.0580 & 49.38 \\
& 1\%   & 39.36 & 0.5823 & 0.0632 & 46.00 & 40.0  & 0.5014 & 0.0534 & 53.44 \\
\midrule
\multirow{3}{*}{\begin{tabular}{c}\textbf{ReCDA+}\\ \textbf{Retrain}\end{tabular}}    
& 0.1\% & 32.81 & 0.5669 & 0.1583 & 4.30  & 44.51 & 0.4878 & 0.2122 & 52.62 \\
& 0.5\% & 33.86 & 0.5480 & 0.1435 & 45.21 & 45.94 & 0.4378 & 0.1447 & 51.22 \\
& 1\%   & 34.96 & 0.5447 & 0.1411 & 45.53 & 54.23 & 0.4270 & 0.1692 & 57.129 \\
\midrule
\multirow{3}{*}{\shortstack[l]{\textbf{NetGuard}\\\textbf{PS}}}   
& 0.1\% & 70.15 & 0.2535 & 0.4359 & 74.65 & 89.07 & 0.1090 & 0.1992 & 89.10 \\
& 0.5\% & 77.17 & 0.1991 & 0.3352 & 80.08 & 94.23 & 0.0596 & 0.1027 & 94.04 \\
& 1\%   & 81.78 & 0.1419 & 0.2686 & 85.81 & 96.19 & 0.0390 & 0.0499 & 96.10 \\
\midrule
\rowcolor{lightgreen}\textbf{NetGuard E2E} & 1\% & 86.00 & 0.1339 & 0.2450 & 88.00 & 98.01 & 0.0256 & 0.0500 & 99.00 \\
\midrule
\textbf{Full} 
& 100\% & 93.22 & 0.0556 & 0.0904 & 93.22 & 98.95 & 0.0102 & 0.0107 & 98.95 \\ 
\bottomrule
\end{tabular}
}
\endgroup
\vspace{-0.2cm}
\end{table}
\vspace{-0.5cm}
\subsection{Baselines}
\vspace{-0.1cm}
We compare \NetGuard with the existing active approaches as well as other concept drift adaptation schemes such as CADE \cite{cade}.For active learning, the first baseline is the most ubiquitously used Uncertainty sampling \cite{uncertainty} that selects samples for which the model produces the most uncertain predictions (where the entropy computed from the softmax layer of a neural network is used for uncertainty prediction). The second baseline includes the Coreset \cite{coreset} approach that selects a subset of samples using K-means clustering approach by choosing each cluster's center as a training point to ensure diversity and coverage in training. The third baseline is CLUE \cite{clue} that applies weighted clustering in the feature space to select instances that are both highly uncertain and diverse to maximize informativeness. For each of the active learning methods, we set the analyst's labeling budget to $0.1\%,\;0.5\%,\;\text{and }1\%$ of the testing dataset. Additionally, we present two constant baselines: (1) a scenario where no adaptation scheme is applied, representing a pure concept drift setting, and (2) a scenario where full adaptation is performed by training on 70\% of the training dataset from the target domain. This evaluation helps determine how \NetGuard performs when no retraining is conducted versus when the classifier is fully retrained using all available training data. In order to gain better insights into the contribution of our end-to-end framework, we also compare it with the state-of-the-art concept drift adaptation technique. Specifically, we utilize CADE \cite{cade} that originally trains a contrastive encoder to identify out-of-distribution (OOD) samples and assigns an OOD score to the samples. We adapt their concept drift adaptation scheme into our prior selection setting in order to select informative samples from the target dataset for retraining.
\vspace{-0.6cm}
\subsection{End-to-end evaluation}
\vspace{-0.1cm}
\subsubsection{CIC 2017 and 2018 Datasets:}
\textbf{Overall adaptation effectiveness.} We evaluate \NetGuard's performance in two key aspects: (1) the effectiveness of the prior selection (PS) module, and (2) the end-to-end performance, which includes both PS and data augmentation using RTF. The results in Table 5 demonstrate the effectiveness of \NetGuard PS and the end-to-end approach (\NetGuard PS + Augmentation) across different labeling budgets and scenarios. \NetGuard PS consistently outperforms other methods, achieving higher F1 scores and lower FPR and FNR. For example, with a 1\% labeling budget, \NetGuard PS achieves an F1 score of 81.78\% for Train: 2017 → Test: 2018 and 96.19\% for Train: 2018 → Test: 2017. Sensitivity experiments on GMM parameters for \NetGuard PS, along with PCA visualizations illustrating its sampling strategy for minority classes, are provided in the Appendix. The end-to-end approach further improves these results, reaching 86.00\% and 98.01\% F1 scores, respectively. This improvement is more pronounced with higher labeling budgets, as more samples enhance the quality of augmented data. It is worth mentioning that \NetGuard PS also benefits from considering the distributional discrepancy between datasets, which is a key factor in its superior performance. In contrast, methods like Uncertainty and CLUE struggle due to their reliance on model performance for sample selection, where the model is already over-fitted to the distorted and limited latent space, leading to the wrong estimation of informativeness for sample selection. For example, the Uncertainty model performs poorly because it tends to misclassify rare classes with low uncertainty (i.e., high confidence), leading to biased sampling towards majority classes. Similarly, CLUE's dependency on model performance and assumption of static latent space, as well as Coreset's distance-based selection, fail to address the distributional discrepancies effectively. In the scenario where training is on 2017 data and testing on 2018 data, the end-to-end approach achieves 86\% F1 with a 1\% labeling budget, close to the Full Adapt scenario that gives an upper bound of 93.22\%. This is because the 2018 dataset is more challenging with more distributional shifts and diverse patterns in contrast with the 2017 dataset. In the reverse scenario (Train: 2018 → Test: 2017), the end-to-end approach nearly matches the Full Adapt performance, highlighting its robustness in handling distributional shifts and achieving near-optimal results.

\noindent \textbf{Class-specific adaptation effectiveness.} The results in Table \ref{tab:combined_micro_f1_scores1} demonstrate the impact of \NetGuard on class-level performance across different attack types. Since certain attack classes exhibited extremely low performance during the prior selection stage of \NetGuard, we applied augmentation specifically to those classes by generating three times the original strength using RTF. For Table \ref{tab:combined_micro_f1_scores1} scenario, those rare attack types include DoS Slowhttptest, FTP-BruteForce, Infiltration, and Web Attack. As shown, the F1-score for DoS Slowhttptest increased from 0.0 with no adaptation to 0.44 using CLUE and 0.63 with \NetGuard PS, further improving to 0.72 with augmentation, exceeding Full Adapt (0.62) due to the added diversity, which enhanced classifier robustness. FTP-BruteForce saw substantial gains, rising from 0.0 to 0.51 with \NetGuard PS and 0.71 with augmentation, closely matching Full Adapt (0.74). Infiltration, one of the most challenging classes due to its rarity, improved from 0.23 with \NetGuard PS to 0.28 after augmentation but still trailed Full Adapt (0.34). Web Attack followed a similar trend. We emphasize a key observation: as we augment the dataset by generating additional samples for Web Attack, FTP-BruteForce, Infiltration, and DoS Slowhttptest and incorporating them into the original CIC-IDS2017 dataset for IDS training, the performance of most other classes remains consistently stable within a range of 0.0 to 0.20. In some cases, augmentation even improves the performance of non-augmented classes, such as DDoS, DoS Hulk, and DoS slowloris, demonstrating that the added diversity helps refine the classifier’s decision boundaries. This finding is significant as the augmentation does not degrade the overall IDS performance, thus validating the feasibility of targeted augmentation in enhancing underrepresented attack detection. We obtained similar improvement for the augmented class in the reverse setting with Train: 2018 → Test: 2017, shown in the Appendix \ref{tab:combined_f1_scores2} 
\vspace{-0.28cm}
\begin{table}[h!]
\centering
\caption{\small NetGuard Class-wise E2E results (Train:2017 \(\rightarrow\) Test:2018)}
\footnotesize
\setlength{\tabcolsep}{1pt}
\renewcommand{\arraystretch}{1}
\begin{tabular}{lc|cccc|c}
\toprule
& & \multicolumn{4}{c}{\textbf{F1 scores (in \%) for 1\% labeling cost}} \vline & \\
\hline
\textbf{Attack Class} & \textbf{No Adapt} & \textbf{CLUE} & \textbf{CADE} & \textbf{ReCDA} & \textbf{NetGuard} & \textbf{Full Adapt} \\
\midrule
Benign              & 0.46 & 0.90 & 0.47 &0.47 & 0.88 & 0.85\\
Bot                 & 0.0  & 0.96 & 0.0 &0.0 & 1.00 & 0 \\
DDoS                & 0.0  & 0.73 & 0.36 &0.0  & 0.92 & 0.99 \\
DoS GoldenEye       & 0.39 & 0.91 & 0.51 &0.41 & 0.96 & 0.99 \\
DoS Hulk            & 0.90 & 0.89 & 0.93 &0.94  & 0.98 & 0.99 \\
\rowcolor{lightgreen} \textbf{DoS Slowhttptest} & \textbf{0.0} & \textbf{0.44} & \textbf{0.00} &\textbf{0.00}  & \textbf{0.72} & \textbf{0.62} \\
DoS slowloris       & 0.92 & 0.85 & 0.97 &0.95  & 0.95 & 1.0 \\
\rowcolor{lightgreen}
\textbf{FTP-BruteForce}   & \textbf{0.0} & \textbf{0.57} & \textbf{0.00} &\textbf{0.00}  & \textbf{0.71} & \textbf{0.74} \\
\rowcolor{lightgreen}
\textbf{Infiltration}     & \textbf{0.0} & \textbf{0.20} & \textbf{0.003} &\textbf{0.15}  & \textbf{0.28} & \textbf{0.34} \\
SSH-Bruteforce      & 0.90 & 1.00 & 0.77 & 0.89  & 1.00 & 1.0 \\
\rowcolor{lightgreen}
\textbf{Web Attack}       & \textbf{0.0} & \textbf{0.24} & \textbf{0.31} &\textbf{0.07} & \textbf{0.50} & \textbf{0.8} \\
\bottomrule
\end{tabular}
\label{tab:combined_micro_f1_scores1}
\end{table}
\vspace{-0.22cm}
.Another key observation is \NetGuard effectiveness in handling \textit{novel attacks}, particularly when dealing with novel attack variants. As seen in Figure 2, new DDoS attack variants emerged in the CIC-IDS2018 (DDoS UDP + DDoS HTTP) dataset compared to CIC-IDS2017 (with only DDoS HTTP). This trend is reflected in Table \ref{tab:combined_micro_f1_scores1}, where, under the Train: 2017 → Test: 2018 setting, DDoS detection without adaptation results in an F1-score of 0.0, indicating a complete failure to recognize the newly introduced attack patterns. However, with classifier's performance significantly improves reaching an F1 score of 0.92 with \NetGuard's end-to-end approach. These results highlight that \NetGuard not only adapts to distribution shifts for similar attack types but is also highly effective in identifying novel attacks and new variants, making it a robust solution for real-world intrusion detection scenarios where emerging threats are a challenge.

\noindent \textbf{Comparison with OOD sample selection using CADE and ReCDA}. To ensure a fair baseline, we adapt CADE~\cite{cade} and ReCDA ~\cite{recda} as sample selection strategies, denoted as CADE+Retrain and ReCDA+ Retrain. Both select target-domain samples based on drift-aware scoring within a fixed labeling budget, combine them with full source data, and retrain an MLP classifier. CADE+Retrain uses a contrastive autoencoder to identify drifted samples via Mahalanobis distance in latent space. It retains the original encoder but retrains the classifier on raw features. Default parameters include: $\tau = 0.5$, $\sigma = 0.7$, $\text{TMAD} = 2$, and $b = 1.4826$. ReCDA+Retrain uses unsupervised scoring by ranking cosine similarity ratios in original vs. embedding space to $K{=}2$ cluster centroids. A shallow encoder trained on source data guides selection. Parameters include $\tau = 0.5$, $\sigma = 0.7$, $\delta$ (selection ratio), and $\epsilon = 10^{-5}$. As shown in Table~\ref{tab:combined_results}, \NetGuard outperforms both (81.78\% F1 vs. 81.35\% for CADE and 72.47\% for ReCDA) using only 1\% labeled samples from 100k data points. 
This improvement arises from significant concept drift between the source and target domains, illustrated in Fig. 1. As illustrated, attack class distributions shift substantially across domains, and embeddings of benign and attack classes begin to overlap in the target domain. These changes violate the core assumptions behind existing methods: CADE assumes intra-class embedding similarity, and ReCDA assumes embedding consistency. CADE achieves low FPR due to contrastive learning enforcing a closed-world boundary—drifted attacks near benign clusters are misclassified as benign. This reduces recall and F1 under drift. \NetGuard handles open-world drift by generating synthetic samples that preserve overlap between classes while maintaining discriminative features. At 1\% labeling, NetGuard achieves superior AUROC scores (0.91/0.99) compared to CADE+Retrain (0.68/0.81) and ReCDA+Retrain (0.67/0.78) for 2017→2018 and 2018→2017 transfers, showing NetGuard’s stronger ability to differentiate between multiple classes—unlike CADE and ReCDA, which lack full multi-class support (CADE experiments are limited to 4 classes in CIC-IDS and ReCDA supports only binary classes). We also experimented with CADE using a relaxed labeling strategy by selecting all drifted samples with parameter TMAD > 2 (22,085 samples, ~22\% labeling cost). Despite this higher labeling cost, CADE achieved only 56.36\% accuracy, 0.794 AUROC, 0.0503 FPR, and 0.4606 FNR, still underperforming compared to \NetGuard with just 1\% labeled data. Prior work shows that false positives increase analyst workload~\cite{sommer}, and labeling capacity is limited to 80 samples/day~\cite{braun}. By using just 1\% labeled data, \NetGuard reduces annotation effort with high accuracy.
\vspace{-0.1in}
\subsubsection{Improved interpretability with NetGuard via Trustee} To interpret the functionality of \NetGuard, we used an interpretable rule-based module, Trustee~\cite{trustee}, which extracts human-readable decision rules by approximating a trained classifier's predictions using compact decision sets. We applied Trustee to analyze rules generated in the CIC-IDS setting (Train on 2017, Test on 2018) under two adaptation strategies: (1) \NetGuard PS with only 1\% labeled samples from the target domain, and (2) Full Adaptation using 100\% of the target data. Our analysis reveals a strong alignment between the two sets of rules. In both cases, key security-relevant features such as \textit{PSH Flag Count}, \textit{Active Mean}, and \textit{Fwd IAT Mean} consistently appear in the decision logic for identifying attacks like Infiltration and Web Attack. These overlaps confirm that \NetGuard’s prior selection (PS) effectively identifies representative and discriminative samples even with minimal supervision. By emphasizing flow-level and timing-based features, \NetGuard maintains interpretability while achieving high detection performance. A detailed comparison of Trustee-derived rules across No Adaptation, \NetGuard PS, and Full Adaptation is provided in the Appendix (Table~\ref{tab:trustee_rules_clean}).
\vspace{-0.1in}
\subsubsection{UGR16:} \textbf{Overall adaptation effectiveness (Train: July \(\rightarrow\) Test: August).} 
 The performance metrics in Table \ref{tab:UGRresults} confirm the presence of concept drift in the UGR'16 dataset, as the No Adapt (0\%) setting causes the F1-score to drop to 68.77\% (MLP) and 74.33\% (XGBoost) when training on July data and testing on August data. With just 1\% labeled data, NetGuard PS + Augmentation achieves 89.66\% F1 (MLP) and 99.29\% F1 (XGBoost), outperforming all baselines and nearing Full Adaptation (93.43\% MLP, 99.55\% XGBoost), proving its efficiency in mitigating drift. 
\vspace{-0.3cm}
\begin{table}[ht!]
\centering
\footnotesize
\caption{\small Performance on UGR'16 (Train: July \(\rightarrow\) Test: August)}
\scriptsize
\setlength{\tabcolsep}{1pt} % Tighten column spacing
\renewcommand{\arraystretch}{0.35} % Tighten row spacing
\label{tab:UGRresults}
\resizebox{0.95\columnwidth}{!}{ % Slightly shrink to fit column
\begin{tabular}{ll|ccc|ccc}
\toprule
& & \multicolumn{3}{c|}{\textbf{MLP}} & \multicolumn{3}{c}{\textbf{XGBoost}} \\
\textbf{Method} & \textbf{Label} & \textbf{F1 (\%)} & \textbf{FNR} & \textbf{FPR} & \textbf{F1 (\%)} & \textbf{FNR} & \textbf{FPR} \\
\midrule
\textbf{No Adapt} & 0\% 
  & 68.77 & 0.300 & 0.08 & 74.33 & 0.246 & 0.062 \\
\midrule
\multirow{3}{*}{\textbf{Uncertainty}} 
  & 0.1\% & 67.68 & 0.2923 & 0.0755 & 79.40 & 0.1855 & 0.0466 \\
  & 0.5\% & 69.02 & 0.2790 & 0.0718 & 85.73 & 0.1403 & 0.0353 \\
  & 1\%   & 68.40 & 0.2864 & 0.0735 & 90.36 & 0.1003 & 0.0253 \\
\midrule
\multirow{3}{*}{\textbf{CLUE}} 
  & 0.1\% & 15.75 & 0.7901 & 0.1981 & 89.04 & 0.0800 & 0.0266 \\
  & 0.5\% & 13.82 & 0.8005 & 0.2002 & 92.31 & 0.0561 & 0.0188 \\
  & 1\%   & 36.91 & 0.5719 & 0.1474 & 94.04 & 0.0446 & 0.0149 \\
\midrule
\rowcolor{lightgreen}
\textbf{NetGuard} & 1\% 
  & 89.66 & 0.0822 & 0.0261 & 99.29 & 0.0065 & 0.0019 \\
\midrule
\textbf{Full Adapt} & 100\% 
  & 93.43 & 0.0500 & 0.0167 & 99.55 & 0.0038 & 0.0012 \\
\bottomrule
\end{tabular}
}
\end{table}
\vspace{-0.2in}
\subsection{Ablation Study}
\begin{figure}[h!] % Single-column figure
    \centering
    \includegraphics[width=0.44\textwidth]{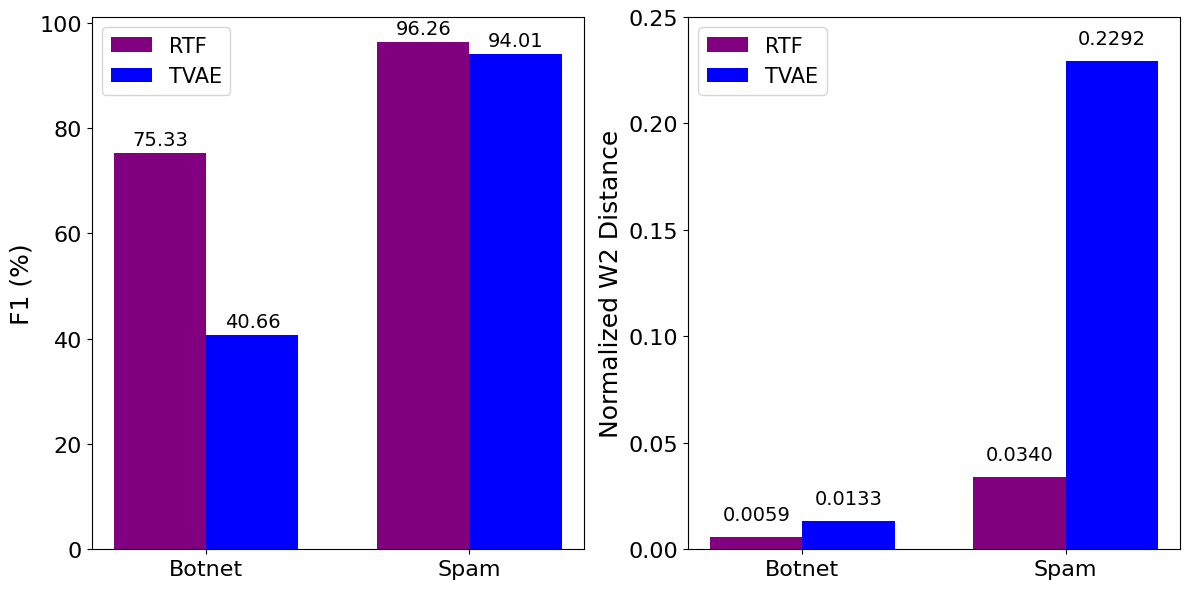} % Adjusted for single-column
    \caption{\small Comparing RTF VS TVAE for UGR16}
    \vspace{-0.1cm}
    \label{fig:tvaertf}
\end{figure}

\vspace{-0.12cm}
We evaluate generative model choices (RTF vs. TVAE) in \NetGuard using UGR'16 (1\% sampling budget, MLP classifier). Comparative analysis includes: (1) augmentation strength (3× vs. 10×) for Spam/ Botnet classes in UGR'16 (see Appendix \ref{fig:ablationappendix}), and (2) synthetic data quality via PCA visualizations (CIC-IDS, Appendix \ref{fig:pcav2}). We compare RTF and TVAE as generative backbones within \NetGuard. As shown in Figure~\ref{fig:tvaertf}, the left graph depicts end-to-end classification performance, where RTF outperforms TVAE in both classification (F1: 75.33\% vs. 40.66\% for Botnet; 96.26\% vs. 94.01\% for Spam) and the right graph shows statistical fidelity (W2 distance as: 0.0059 vs. 0.0133 for Botnet; 0.0340 vs. 0.2292 for Spam). Our W2 scores surpass NetShare's reported EMD values for UGR16 dataset ($\approx$0.10--0.15) \cite{netshare}. For CIC-IDS 2018, our normalized W2 distances between synthetic and original samples/class are 0.34 (Infiltration), 0.92 (Web Attack), 0.03 (FTP-BruteForce)—reflect class-specific fidelity, comparable to NetShare’s reported EMD values.
\vspace{-0.13in}
\vspace{-0.45cm}
\section{Conclusion}
\vspace{-0.2cm}
\NetGuard is a novel generative framework for NIDS that combines distribution-shift aware sample selection with conditional data augmentation to address concept drift and class imbalance. Our method conditions synthetic sample generation on drift patterns to better adapt to evolving threats, particularly rare attacks like web attack and botnet. Evaluations on CIC and UGR'16 datasets demonstrate improved detection in dynamic environments while reducing adaptation costs. The framework's generative approach maintains data diversity and enhances robustness against both common and emerging attacks.
% \section{Conclusion}

% We proposed a framework that combines distribution-aware active learning with deep generative model-based augmentation to address concept drift in NIDS. Our approach improves model adaptability to evolving threats, particularly for rare attack classes like Botnet. Evaluation on CIC and UGR’16 NIDS datasets demonstrated significant performance gains, with active learning enhancing sample selection and deep generative models improving data diversity. The integration of these techniques enables efficient, scalable, and adaptive NIDS, reducing reliance on costly full adaptation while improving detection in dynamic network environments.

\newpage
\bibliographystyle{ACM-Reference-Format}
\bibliography{main}

\clearpage
\appendix
\section{Appendix}
\begin{figure}[!t] % Single-column figure
    \centering
\includegraphics[width=0.35\textwidth]{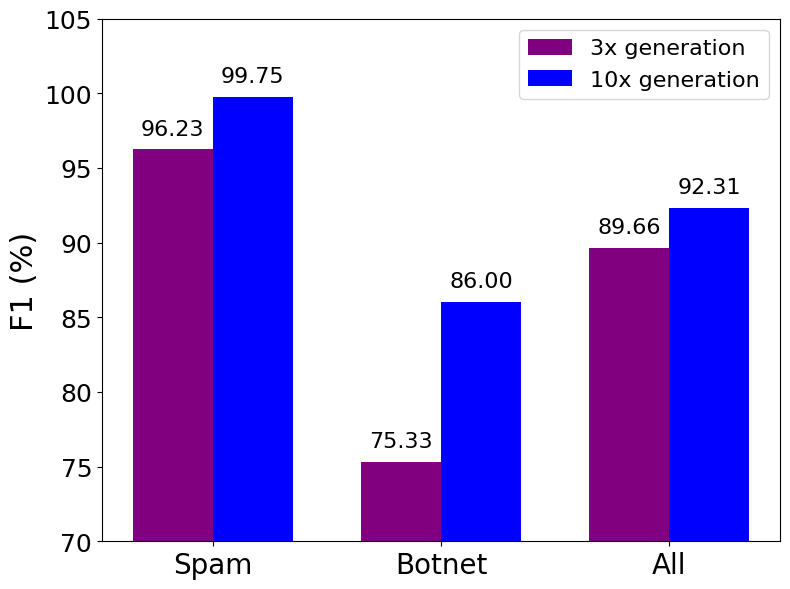} % Adjusted for single-column
    \caption{Impact of generation strength on UGR16}
    \label{fig:ablationappendix}
\end{figure}

\paragraph{Effects of data generation amount on UGR'16}
Fig. \ref{fig:ablationappendix} compares the F1-score for Spam, Botnet, and overall macro F1 when using 3× and 10× data augmentation. As shown, As shown, increasing the volume by 10× improves Spam F1 (96.23\% → 99.75\%), Botnet F1 (75.33\% → 86.00\%), and overall F1 (89.66\% → 92.31\%), demonstrating that generating more samples for training enhances \NetGuard end-to-end classification performance. Beyond 10× augmentation, improvements were minimal, indicating diminishing returns, where initial increases in synthetic data yield significant gains, but further augmentation does not help due to sample redundancy and model over-fitting.
\paragraph{Class-specific adaptation effectiveness for UGR'16 dataset.}
Table \ref{tab:mlp_xgboost_comparison} shows class-specific performance for MLP and XGBoost under different adaptation methods. Spam and Botnet, which had the lowest counts in the original dataset, showed minimal improvement with active learning alone. However, after 3× augmentation, their performance significantly improved. For MLP, Botnet rose from 36.20\% (No-Adapt) to 75.33\% (NetGuard) and Spam from 80.91\% to 96.26\%, approaching Full Adaptation. Similarly, for XGBoost, Botnet improved from 19.39\% to 98.54\%, and Spam from 81.36\% to 99.74\%, demonstrating that augmentation effectively enhances rare class detection across both models.

\begin{table}[!ht]
\centering
\small % Reduce font size
\setlength{\tabcolsep}{3pt} % Reduce column spacing
\renewcommand{\arraystretch}{1} % Adjust row spacing slightly
\caption{\small Class-wise performance of \textbf{MLP} and \textbf{XGBoost} for adaptation methods on UGR'16 (Train: July \(\rightarrow\) Test: August)}
\label{tab:mlp_xgboost_comparison}
\resizebox{\columnwidth}{!}{%
\begin{tabular}{c|ccc|ccc}
\hline
\multirow{2}{*}{\textbf{Attack}} 
& \multicolumn{3}{c|}{\textbf{MLP}} 
& \multicolumn{3}{c}{\textbf{XGBoost}} \\
\cline{2-7}
& \textbf{No-Adapt} & \textbf{NetGuard} & \textbf{Full Adapt} 
& \textbf{No-Adapt} & \textbf{NetGuard} & \textbf{Full Adapt} \\
\hline
Benign  & 68.53  & 83.65  & 89.32  & 73.22  & 98.00  & 99.25 \\
DoS     & 98.53  & 99.02  & 99.78  & 99.72  & 1.00   & 1.00  \\
\textbf{Botnet}  & \textbf{36.20}  & \textbf{75.33}  & \textbf{85.42}  
                 & \textbf{19.39}  & \textbf{98.54}  & \textbf{99.00} \\
Scan44  & 95.67  & 96.76  & 97.30  & 98.49  & 99.68  & 99.81 \\
\textbf{Spam}    & \textbf{80.91}  & \textbf{96.26}  & \textbf{98.30}  
                 & \textbf{81.36}  & \textbf{99.74}  & \textbf{1.00}  \\
\hline
\end{tabular}
}
\end{table}

\paragraph{Class-specific adaption effectiveness in CIC Train: 2018 → Test: 2017 setting.} In the reverse case for CIC experiments as shown in Table \ref{tab:combined_f1_scores2}, augmentation was applied only to Infiltration, as it had the lowest classification performance after the \NetGuard PS stage. The results show that Infiltration's F1-score improved from 0.02 with \NetGuard PS to 0.39 with augmentation, nearly matching Full Adapt (0.40).
This overall validates that \NetGuard PS combined with augmentation reduces the dependency on fully labeled datasets by bringing model performance closer to the Full Adaptation scenario. 

\begin{table}[h!]
\centering
\small
\caption{\small Class-level performance for Train: 2018\(\rightarrow\) Test: 2017}
\scriptsize % Reduce font size for better fit
\setlength{\tabcolsep}{0.6pt} % Adjust column spacing for compact fit
\renewcommand{\arraystretch}{0.6} % Adjust row spacing
\begin{tabular}{lcccccc}
\toprule
\textbf{Attack} & \textbf{No Adapt} & \textbf{1\% CLUE} & \textbf{1\% NetGuard PS} & \multicolumn{1}{c}{\textbf{End-to-end}} & \textbf{Full Adapt} \\
                &                   &                 &                    & \multicolumn{1}{c}{\textbf{(NetGuard PS+Data Aug.)}} & \\
\midrule
Benign              & 0.46 & 0.93 & 0.95 & 0.94 & 0.99 \\
Bot                 & 0.0  & 0.00 & 0.73 & 0.75 & 0.91 \\
DDoS                & 0.54 & 0.93 & 0.97 & 0.97 & 1.0  \\
DoS GoldenEye       & 0.77 & 0.81 & 0.96 & 0.96 & 0.99 \\
DoS Hulk            & 0.20 & 0.91 & 0.95 & 0.95 & 0.99 \\
DoS Slowhttptest    & 0.0  & 0.40 & 0.97 & 0.96 & 0.99 \\
DoS slowloris       & 0.6  & 0.76 & 0.96 & 0.97 & 0.99 \\
FTP-BruteForce      & 0.0  & 0.91 & 0.99 & 0.98 & 1.0  \\
\textbf{Infiltration} & \textbf{0.0}  & \textbf{0.00}  & \textbf{0.02}  & \textbf{0.39}  & \textbf{0.40} \\
SSH-Bruteforce      & 0.9  & 0.93 & 0.98 & 0.97 & 0.99 \\
Web Attack          & 0.07 & 0.88 & 0.90 & 0.90 & 0.94 \\
\bottomrule
\end{tabular}
\label{tab:combined_f1_scores2}
\end{table}

\paragraph{Sensitivity analysis for GMM-based prior selection.}

Table 8 illustrates that using 1\% labeling, a GMM with K=10 and diagonal covariance achieves the best trade-off between F1-score and runtime across both CIC-IDS 2017/18 and UGR'16, making it the optimal setting for prior selection under drift; its parametric nature enables efficient, scalable modeling of distributional shifts, unlike non-parametric methods which are often computationally expensive and sensitive to hyperparameter tuning. 
\vspace{-0.15cm}
\begin{table}[ht]
\centering
\footnotesize         
\caption{\small Sensitivity of GMM parameters at 1\% labeling}
\label{tab:gmm-sensitivity}
\setlength{\tabcolsep}{7pt}
\begin{tabular}{@{}lcccr@{}}
\toprule
\textbf{Dataset}      & \textbf{K} & \textbf{Covariance} & \textbf{F1} & \textbf{Runtime (s)} \\ \midrule
\multirow{3}{*}{CIC-IDS 2017/18} %
    & 5  & diagonal & 0.81 & 3.48  \\
    & \textbf{10} & diagonal & \textbf{0.84} & 5.97  \\
    & 10 & full     & 0.83 & 12.96 \\ \midrule
 UGR'16 (Train: July \(\rightarrow\) Test: Aug.)               & \textbf{10} & diagonal & \textbf{0.71} & 6.32  \\ \bottomrule
\end{tabular}
\end{table}
\vspace{-0.3cm}
\paragraph{Computational cost of NetGuard.} Table 9 presents a breakdown of the computational usage of various {\tt NetGuard} components. {\tt NetGuard}'s REaLTabFormer training takes $\approx$40 minutes per cycle on an A100 GPU (avg. 25\% utilization), while sample generation costs 4–6 minutes per 1,000 samples (0.24–0.36s/sample, <5\% GPU). GMM-based prior selection takes 6 seconds per 1,000 samples ($\approx$6ms/sample, CPU-bound). 
\begin{table}[ht]
\centering
\caption{\small Computational Overhead of {\tt NetGuard}}
\label{tab:netguard-costs}
\footnotesize
\setlength{\tabcolsep}{1pt}
\renewcommand{\arraystretch}{1.00}
\begin{tabular}{@{}lccr@{}}
\toprule
\textbf{Component} & \textbf{Cost / 1k Samples} & \textbf{Latency / Sample} & \textbf{Hardware Util.} \\
\midrule
GMM-based prior selection & 5.97–6.32 sec & $\approx$ 6 ms & CPU-bound / <5\% \\
RTF training (initialization) & ~40 min & — & GPU / ~25\% \\
RTF sample generation & 4–6 min & 0.24–0.36 sec & GPU / <5\% \\
\bottomrule
\end{tabular}
\end{table}
\paragraph{PCA visualization of attack classes before and after data augmentation}
We provide PCA visualizations of CIC-IDS 2018 attack samples before (left) and after (right) 3× data augmentation in Fig.5. The augmented samples closely follow the original distribution while remaining distinguishable across classes.
\begin{figure}[!ht]
    \centering
    \includegraphics[width=\linewidth]{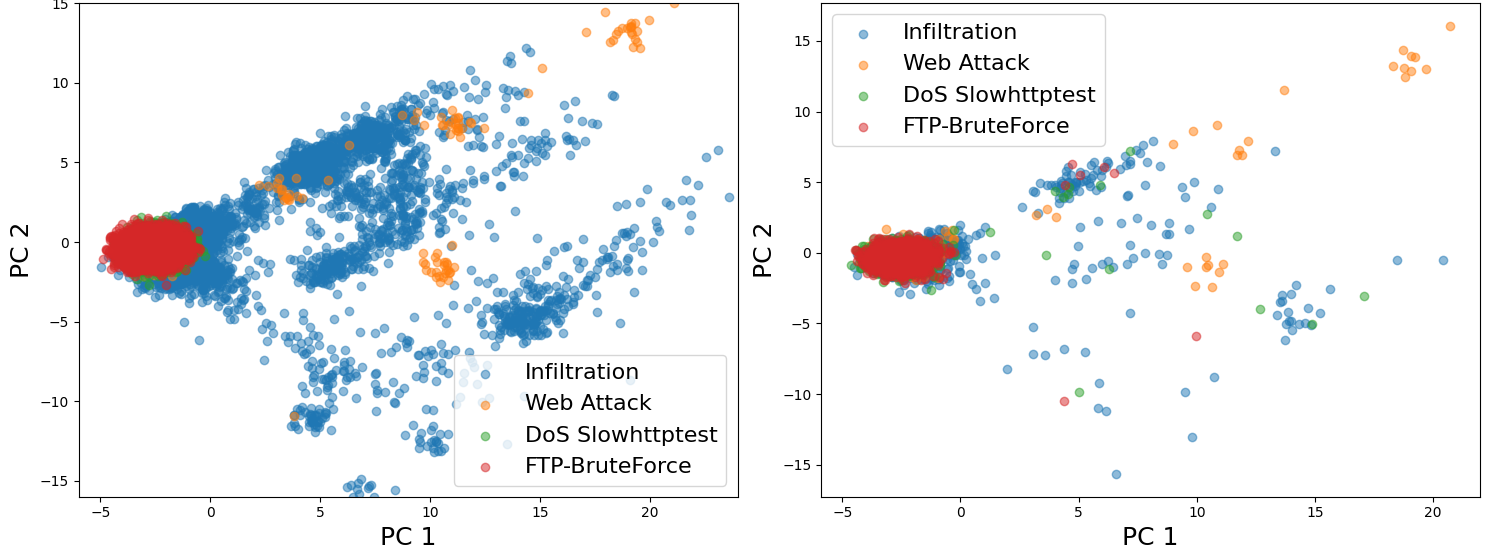}
    \caption{\small PCA projection of four CIC-IDS 2018 attack classes before (all samples, left) and after (3× augmentation on only the PS samples, right). }
        \label{fig:pcav2}
\end{figure}

\begin{figure*}[!ht]
    \centering
    %----------- Subfigure (a) -----------
    \begin{subfigure}[b]{0.3\textwidth}
        \includegraphics[width=\textwidth]{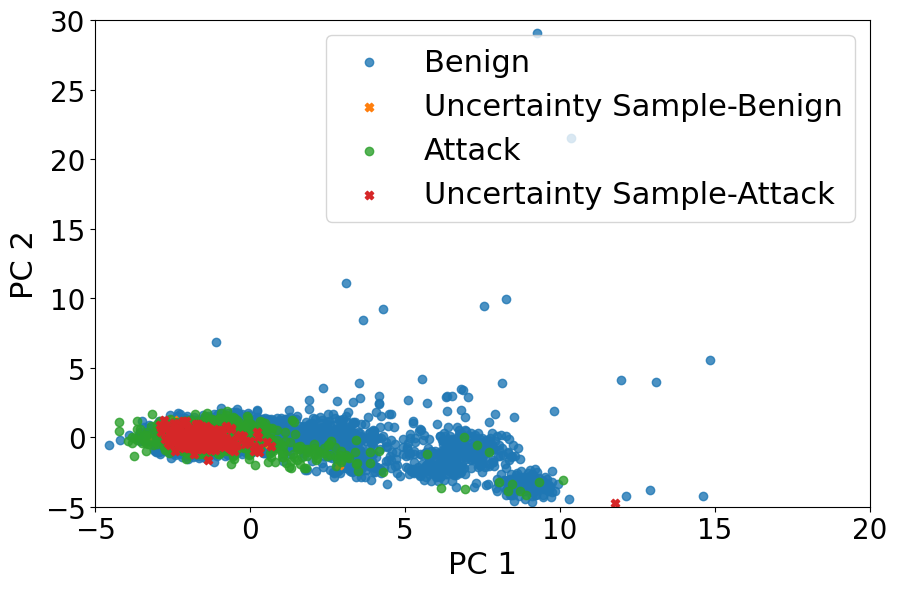}
        \caption{Uncertainty Sampling}
        \label{fig:subfig1}
    \end{subfigure}
    \hfill
    %----------- Subfigure (b) -----------
    \begin{subfigure}[b]{0.3\textwidth}
        \includegraphics[width=\textwidth]{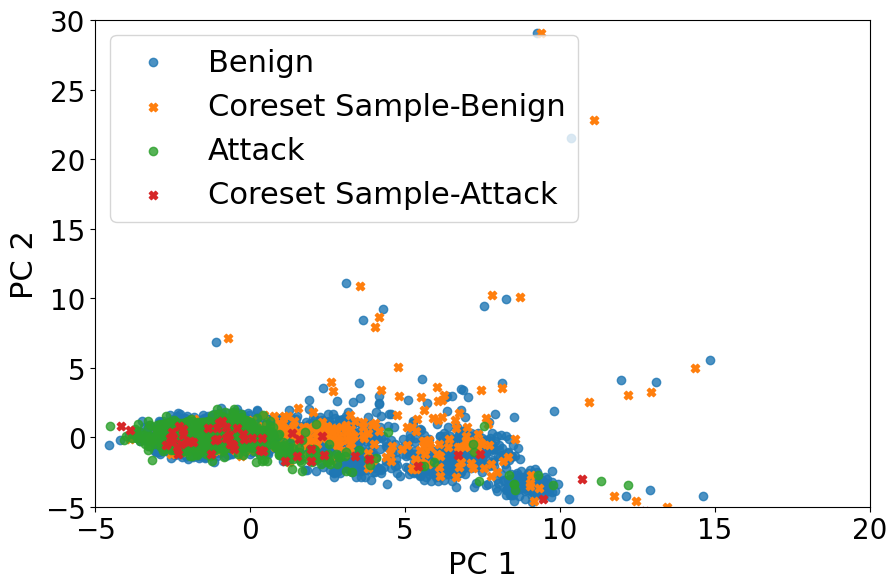}
        \caption{Coreset Sampling}
        \label{fig:subfig2}
    \end{subfigure}
    \hfill
    %----------- Subfigure (c) -----------
    \begin{subfigure}[b]{0.3\textwidth}
        \includegraphics[width=\textwidth]{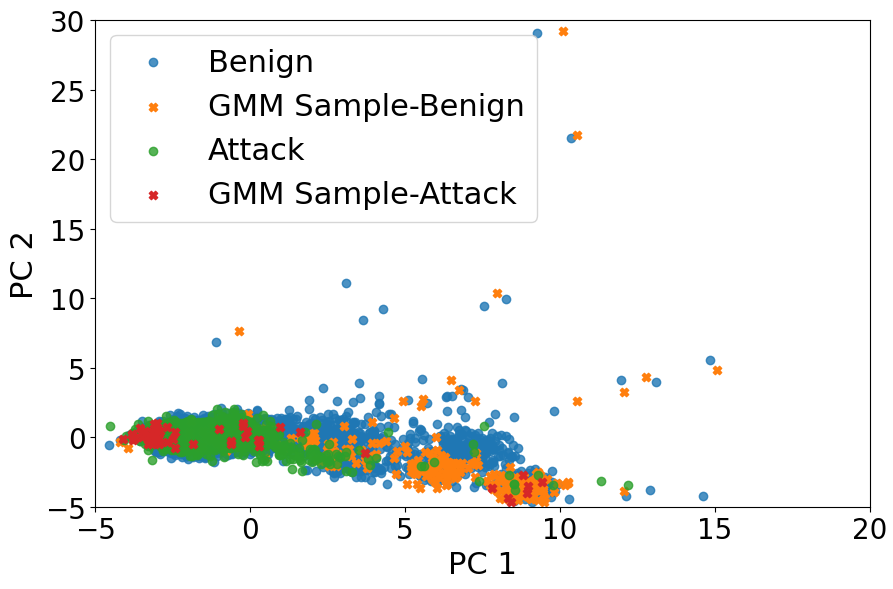}
        \caption{Density-aware Sampling (Ours)}
        \label{fig:subfig3}
    \end{subfigure}

    \caption{%
    Comparison of latent space visualizations for different prior selection sampling strategies
    }
    \label{fig:samplingpca}
\end{figure*}

\paragraph{Evolving trustee rules on CIC 2017-> 2018 dataset} Table \ref{tab:trustee_rules_clean} shows the evolving Trustee classification rules for Infiltration, Web Attack, and FTP-BruteForce under different adaptation scenarios on CIC-IDS 2017→2018. Our analysis of classifier decision rules (Table 10) demonstrates that \NetGuard's prior selection (1\% labeling) achieves 89\% feature commonality with full target domain adaptation (100\%), validating its ability to identify the most critical drift indicators (e.g., Fwd IAT Mean thresholds, PSH flag counts) with minimal supervision. The detailed rule breakdown in the Appendix reveals consistent preservation of core detection logic across all attack classes - Infiltration maintains flow activity thresholds, Web Attacks adapt packet timing features while preserving structural patterns, and FTP-BruteForce retains segment size requirements - confirming that \NetGuard's efficient approach captures nearly all discriminative features of expensive full adaptation. This strong alignment between limited and full adaptation rules proves \NetGuard's effectiveness for practical deployment where exhaustive retraining is infeasible.
\paragraph{PCA visualization of CIC 2018 embedding space with different sampling methods.} PCA visualization of different sampling approaches on CIC 2017 dataset (for binary classes; attack and benign).
Figure \ref{fig:samplingpca}  illustrates the limitations of traditional sampling strategies using PCA visualizations. Subfigures (a) and (b) demonstrate that uncertainty sampling and CoreSet sampling tend to focus on high-density regions, often ignoring minority attack samples. In contrast, our density-aware approach (Figure \ref{fig:samplingpca}) prioritizes low-density regions, ensuring that critical attack types are better represented in the training set.
\begin{table}[t]
\scriptsize
\renewcommand{\arraystretch}{1.15}
\setlength{\tabcolsep}{2pt}
\centering
\caption{\small Trustee classification rules (Train:2017 \(\rightarrow\) Test:2018)}
\begin{tabular}{p{1.2cm} p{1cm} p{2.8cm} p{2.8cm}}
\toprule
\textbf{Class} & \textbf{No Adaptation} & \textbf{\NetGuard PS (1\%)} & \textbf{Full Adaptation (100\%)} \\
\midrule
\textbf{Infiltration} &
--- &
PSH Flag Count > 0.16 \newline
Active Mean > 20.76 &
Dest Port $\leq$ 0.30, Bwd Header Len $\leq$ 0.01, Flow IAT Mean $\leq$ 0.33, min\_seg\_size\_fwd $\leq$ 0.00 \newline
Active Min > 3.64, Fwd IAT Max $\leq$ 0.23 \\
\textbf{Web Attack} &
--- &
\textcolor{blue}{Fwd IAT Mean > 0.33}, Fwd IAT Total $\leq$ 0.53, Fwd IAT Min $\leq$ 0.18 \newline
Flow Packets/s > 0.27, Flow Duration $\leq$ 0.56 &
Init Win Bytes Fwd > 0.72, \textcolor{blue}{Fwd IAT Mean $\leq$ 0.29}, Flow Bytes/s $\leq$ 0.05 \newline
Fwd Packet Len Std > 1.37 \\
\textbf{FTP-BruteForce} &
Bwd Packets/s &
\textcolor{blue}{min\_seg\_size\_fwd > 0.00}, Flow Duration $\leq$ 0.56, Packet Len Var $\leq$ 0.39 \newline
Flow IAT Max > 0.49, Flow IAT Mean $\leq$ 0.33 &
\textcolor{blue}{min\_seg\_size\_fwd > 0.00}\newline
Fwd Packets/s > 0.16 \\
\bottomrule
\end{tabular}
\label{tab:trustee_rules_clean}
\end{table}

\end{document}